 \documentclass[runningheads]{llncs}
\usepackage{times}

\usepackage{amsmath, amssymb}

\usepackage{graphicx}
\usepackage{wrapfig}

\newcommand{\poly}{\mathrm{poly}}

\newcommand{\keywords}[1]{\par\addvspace\baselineskip
\noindent\keywordname\enspace\ignorespaces#1}

\title{Quasiperiodicity and non-computability in tilings}

\author{Bruno Durand\inst{1}%
\and Andrei Romashchenko\inst{1,2}}

\authorrunning{B. Durand and A. Romashchenko}
\institute{%Laboratoire d'Informatique, de Robotique et de Microelectronique de Montpellier 
LIRMM, Univ. Montpellier \& CNRS
\and
On leave from IITP RAS
}
\begin{document}

\maketitle

\begin{abstract}
We study tilings of the plane that combine strong properties of  different nature: combinatorial and algorithmic.  
We  prove the existence of a tile set that accepts only quasiperiodic and non-recursive tilings. Our construction is based on the fixed point construction~\cite{drs}; we improve this general technique  and make it enforce the property of local regularity of tilings  needed for quasiperiodicity. 
We prove also a stronger result:  any  $\mathrm \Pi_1^0$-class can be recursively transformed into a tile set %(defining by itself another $\Pi_1^0$-set) 
so that the Turing degrees of the resulting tilings consists exactly of the upper cone based on the Turing degrees of the latter.

 \keywords{tiling, Wang tiles, computability, quasiperiodicity,  fixed point}
\end{abstract}

\section{Introduction}

Tilings form a popular basis for many mathematical games, for games for the kids. In science, they are popular tools for rather different researches, in chemistry (to describe quasicrystalline structures, e.g., \cite{levitov}), in pure logics (e.g. deciding classes of first order predicates defined on their syntax, see \cite{gurevitch}), in computational complexity (as basic model for complexity, \cite{boas}). The first famous result about tilings is the so-called domino problem: Berger proved that given a tile set, we cannot decide algorithmically whether  it can tile the plane,  \cite{berger}. Within the proof, Berger constructed the first aperiodic tile set --- a tile set that can tile the plane but only non-periodically.  It was  the first  tile set that allows only  tilings of the plane with rather complex structure.  Thus, rather simple local rules can imply quite nontrivial global structure of a tiling.

Since Berger's paper, quite a lot of different algorithmic and combinatorial properties of aperiodic tilings were investigated. It was proven that  a tile set that accepts only aperiodic tilings, must accept uncountably many of them, \cite{bruno}. Many researchers  tried to construct  possibly simpler aperiodic tile sets (e.g., \cite{robinson}, \cite{gs}, \cite{kari}, \cite{ollinger}, \cite{drs}). The idea of ``simplicity'' was interpreted in several different ways: as the number of tiles, algorithmic simplicity of the construction, etc. Another avenue of research was constructing tile sets that guarantee not only aperiodicity, but also  more sophisticated  properties  of tilings: non-recursivity, maximal algorithmic complexity (of each tiling), robustness and fault-tolerance of tilings, and their combinations, \cite{nonrecursive1}, \cite{nonrecursive2}, \cite{dls}, \cite{drs}.

The fundamental question ``\emph{How complex can a tiling be?}'' also can be understood in terms of Turing degrees of unsolvability.   Some partial answers to this question are known.  First of all, we remark that for each tile set, the set of valid tilings is effectively closed  (i.e., belongs to the class $\mathrm\Pi_1^0$). 
In \cite{dls} the property of cone-avoidance  was proven: for each tile set $\tau$ and for every undecidable set $A$ there exists a $\tau$-tiling $T$ such that $A$ is not Turing-reducible to $T$.
Quite a complete study of Turing degrees of tilings was given in \cite{pascal1} and \cite{pascal2}. 

Not surprisingly, the constructions that guarantee some nontrivial combinatorial properties or involve simulation of a Turing machine require very different technical features. So it is  rather difficult to combine in one and  the same tiling properties of different nature.
In this paper we try to do some kind of aggregation;  we combine  the combinatorial property of quasiperiodicity with complexity issues. We prove that all upper cones of Turing degrees above any $\mathrm \Pi_1^0$ class can be achieved by a tile set that produces only quasiperiodic tilings. This rather complex theorem has a more concrete consequence: we build a tile set that produces only quasiperiodic tilings, and none of these tilings is recursive.

Let us be more precise now. In this paper  \emph{Wang tiles} are unit squares with colored sides. A \emph{tile set} is a finite family of tiles. For a given tile set the domino problem is to decide whether the entire plane can be tiled with these tiles. Here we assume of course that we are given infinitely many copies of each tile (tiles are prototypes); in other words, we are allowed to  place translated copies of the same tile into different sites of the plane (rotations are not allowed). In a correct tiling the tiles in the neighbor cells must match (sides in contact must have the same color). 

If a tile set $\tau$ tiles the plane, we call these tilings $\tau$-tilings. More formally,  a $\tau$-tiling can be defined as a mapping $F\colon\mathbb{Z}^2\mapsto \tau$, where for each pair of neighboring cells  $x,y\in \mathbb{Z}^2$  the colors of the tiles $F(x)$ and $F(y)$  match each other on their neighboring sides.
A tiling is called \emph{periodic} if some nontrivial shift transforms it into itself. A tiling $F$ is called \emph{quasiperiodic} (or \emph{uniformly recurrent}) if 
every pattern that appears in this tiling,  appears in every sufficiently large square of~$F$.

The domino problem (existence of a tiling with a given tile set) is algorithmically undecidable, \cite{berger}.  An interesting and nontrivial fact (which follows from Berger's theorem) is that there exist tile sets that allow only aperiodic tilings of the plane.

The main result of this article is the following theorem that claims that some tile sets enforce at once two nontrivial properties of a tiling: 
quasiperiodicity and non-computability.
\begin{theorem}\label{thm1}
There exists a tile set \textup(a set of Wang tiles\textup) $\tau$ such that 
%\begin{enumerate}
%\item[(i)] there exist $\tau$-tilings of the plane,
%\item[(ii)] all $\tau$-tilings are quasiperiodic,
%\item[(iii)] all $\tau$-tiling are non-computable.
%\end{enumerate}
(i) there exist $\tau$-tilings of the plane,
(ii) all $\tau$-tilings are quasiperiodic,
(iii) all $\tau$-tiling are non-computable.
\end{theorem}
The tile set from Theorem~\ref{thm1} is \emph{not minimal} (we cannot claim that all $\tau$-tilings contain the same finite pattern). In fact, minimality cannot be combined with non computability: each minimal tile set allows at least one computable tiling, see \cite{ballier-jeandel}.
On the other hand, minimality can be combined with aperiodicity, see Theorem~\ref{thm2} below.

%??With  essentially the same technique we prove a more general  result:
\begin{theorem}\label{thm1-bis}
For every effectively closed set  $\cal A$ there exists a tile set $\tau$ such that 
%\begin{enumerate}
%\item[(i)] all $\tau$-tilings are quasiperiodic,
%\item[(ii)] the Turing degrees of all $\tau$-tiling make up exactly the upper cone of  ${\cal A}$ (defined as  the set of all Turing degrees $d$ such that $d\ge_T \omega $ for at least one $\omega\in {\cal A}$).
%\end{enumerate}
(i) all $\tau$-tilings are quasiperiodic,
(ii) the Turing degrees of all $\tau$-tiling make up exactly the upper cone of  ${\cal A}$ (i.e.,  the class of all Turing degrees $d$ such that $d\ge_T \omega $ for at least one $\omega\in {\cal A}$).

\end{theorem}
For every tile set $\tau$, the set of $\tau$-tilings is alway effectively closed. Moreover, if all $\tau$-tilings are strongly quasiperiodic, then the class of Turing degrees of all $\tau$-tilings is known to be upward closed, see \cite{pascal2}. Thus, Theorem~\ref{thm1-bis} gives a precise characterisation of the Turing spectra of quasiperiodic tilings: they are exactly the upward closed sets in $\mathrm \Pi_1^0$.
Notice that Theorem~\ref{thm1-bis} does not imply the result of \cite{pascal3} (a construction of a minimal subshift of finite type with a nontrivial Turing spectrum that consists of uncountably many cones with disjoint bases). The reason is again that the tile sets constructed in Theorem~\ref{thm1-bis} are not minimal.

We prove Theorem~\ref{thm1} and Theorem~\ref{thm1-bis} using the technique of fixed-point tilings from~\cite{drs}, with some suitable extensions. Though conceptually this technique is  not very difficult,  a very formal explanation would involve many (sometimes excessive) technical  details. 
In order to meet the space limitations of the conference proceedings and also to make the argument more accessible, we present it in a less formalised way, starting with a proof of a simpler Theorem~\ref{thm2} below. Being somewhat sketchy, we nevertheless do not skip any important part of the construction, and  we emphasise the parallels and differences with the previously known construction of a  fixed-point tilings in~\cite{drs}.

The rest of the paper is organised as follows. First we remind the reader the core ideas of the fixed-point tiling from \cite{drs} and explain how this technique implies aperiodicity of tilings. Then we upgrade the construction and build a tile set that combines the properties of aperiodicity and quasiperiodicity. After that we prove the main results of the paper.

\section{Self-simulating tilings (reminder)}
Our proof is based on the \emph{fixed point} construction from \cite{drs}. The main idea of this argument is that we can enforce in a tiling a kind of   \emph{self-similar} structure.  In what follows we remind the principal ingredients of this construction (here we follow the notations from \cite{drs}).
The reader familiar with the technique used in  \cite{drs} can skip this section and go directly to Section~3.

 Let $\tau$ be a tile set and $N>1$ be an integer. We call by a \emph{macro-tile} an $N \times N $ square correctly tiled by matching tiles from $\tau$. Every side of a $\tau$-macro-tile contains  a sequence of $N$ colors (of tiles from $\tau$); we refer to this sequence as a \emph{macro-color}.
Further, let $\rho$ be some set of $\tau$-macro-tiles (of size $N\times N$). We say that $\tau$ \emph{implements} $\rho$ if (i) some $\tau$-tilings exist, and (ii) for every $\tau$-tiling there exists a unique lattice of vertical and horizontal lines that cuts this tiling into $N\times N$ macro-tiles from $\rho$. (We do not require that all macro-tiles from $\rho$ appear in every $\tau$-tiling.) The value of $N$ is called the  \emph{zoom factor} of this implementation.
  
A  tile set $\tau$ is called \emph{self-similar} if it implements some set $\rho$ of $\tau$-macro-tiles with some zoom factor $N>1$ and $\rho$ is isomorphic to $\tau$. 
This means that there exist a one-to-one correspondence between $\tau$ and $\rho$ such that
the matching pairs of $\tau$-tiles correspond exactly to the matching pairs of $\rho$-macro-tiles.
By definition, for a  self-similar tile set $\tau$ each tiling can be uniquely split into $N\times N$ macro-tiles (the set of all macro-tiles is isomorphic to the initial tile set $\tau$);  further, the grid of macro-tiles can be grouped into blocks of size $N^2\times N^2$, where each block is a macro-tile of rank $2$ (again, the set of all macro-tiles of rank $2$ is isomorphic to the initial tile set $\tau$), etc. It is not hard to deduce from this observation the following statement.
\begin{proposition}[folklore] \label{thm-folklore}
A self-similar tile set $\tau$ has only aperiodic tilings.
 \end{proposition}
The proof is based on a simple observation: every period (if it exists) should be a multiple of $N$, since  the lattice of vertical and horizontal lines that cuts this tiling into $N\times N$ macro-tiles must be \emph{unique}. Similarly, a period must be a  multiple of $N^2$ (to respect the uniquely defined grid of macro-macro-tiles), a  multiple of $N^3$, etc. It follows that a period must greater than any integer; see details in \cite{drs}.
 Thus, if we want to construct an aperiodic tile set, then it is enough to present an instance of a self-similar tile set.
 Below we discuss a very general construction of self-similar tile sets.
 
 \subsection{Implementing some given tile set with a large enough zoom factor}
 %%%%
 
 Assume that we have a tile set $\rho$ where each color is a $k$-bit string (i.e., the set of colors $C \subset \{0,1\}^k$) and the set of tiles $\rho \subset C^4$ is presented by a predicate $P(c_1,c_2,c_3,c_4)$ (the predicate is true if and only if the quadruple $(c_1,c_2,c_3,c_4)$ corresponds to a tile from $\rho$). Assume that we have some Turing machine $\cal M$ that computes $P$. Let us show how to implement $\rho$ using some other tile set $\tau$, with a large enough zoom factor $N$.
 
We will build a tile set $\tau$ where each tile ``knows'' its coordinates modulo $N$. This information is included in the tiles' colors. More precisely, for a tile that is supposed to have  coordinates $(i,j)$ modulo $N$, the colors on the left and on the bottom sides should involve $(i,j)$, the color on the right side should involve $(i+1\mod N, j)$, and the color on the top side, respectively, involves $(i, j+1 \mod N)$, see Fig.~1. 
\begin{figure}
\center
\begin{minipage}[b]{0.45\linewidth}
\center
\includegraphics[scale=0.5]{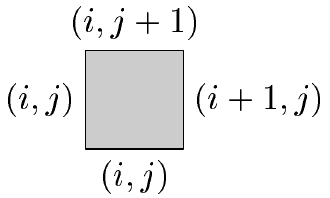}
\label{pic1}
\caption{}
\end{minipage}
\begin{minipage}[b]{0.40\linewidth}
\center
\includegraphics[scale=0.75]{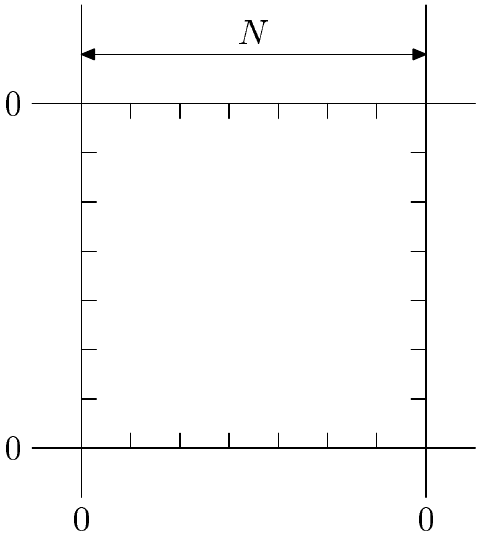}
\label{pic2}
\caption{}
\end{minipage}
\end{figure}
This means that every $\tau$-tiling can be uniquely split into blocks (macro-tiles) of size $N\times N$, where the coordinates of cells ranges from $(0,0)$ in the bottom-left corner to $(N-1,N-1)$ in top-right corner, Fig.~2. So,  intuitively, each tile ``knows'' its position in the corresponding macro-tile.

 In addition to the coordinates, each tile in $\tau$ should have some supplementary information   encoded in the colors on its sides. We refer to this additional information as the \emph{shade} of the color. On the border of a macro-tile (where one of the coordinates is zero) only two additional shades (say, $0$ and $1$) are allowed. Thus, for each macro-tile of size $N\times N$ the corresponding macro-colors represent a string of $N$ zeros and ones. We will assume that $k\ll N$.  We allocate  $k$ bits in the middle of a macro-tile sides and make them represent colors from $C$; all other bits on the sides of a macro-tile are zeros.
  
 Now we introduce additional restrictions on tiles in $\tau$ that will guarantee the required property:  the macro-colors on the macro-tiles satisfy the relation $P$. To achieve this, we ensure that bits from the macro-tile side are transferred to the central part of the tile, and the central part of a macro-tile is used to simulate  a  computation of the predicate $P$.
We fix which cells in a macro-tile are ``wires'' (we may assume that wires do not cross each other) and then require that these tiles carry the same (transferred) bit on two sides. The central part of a macro-tile (of size, say $m\times m$) should represent a time-space diagram of $\cal M$'s computation (the tape is horizontal, time goes up). This is done in a standard way. We require that computation terminates in an accepting state (if not, no correct tiling can be formed), see Fig.~3.
\begin{figure}
\center
\begin{minipage}[b]{0.45\linewidth}
\center
\includegraphics[scale=0.43]{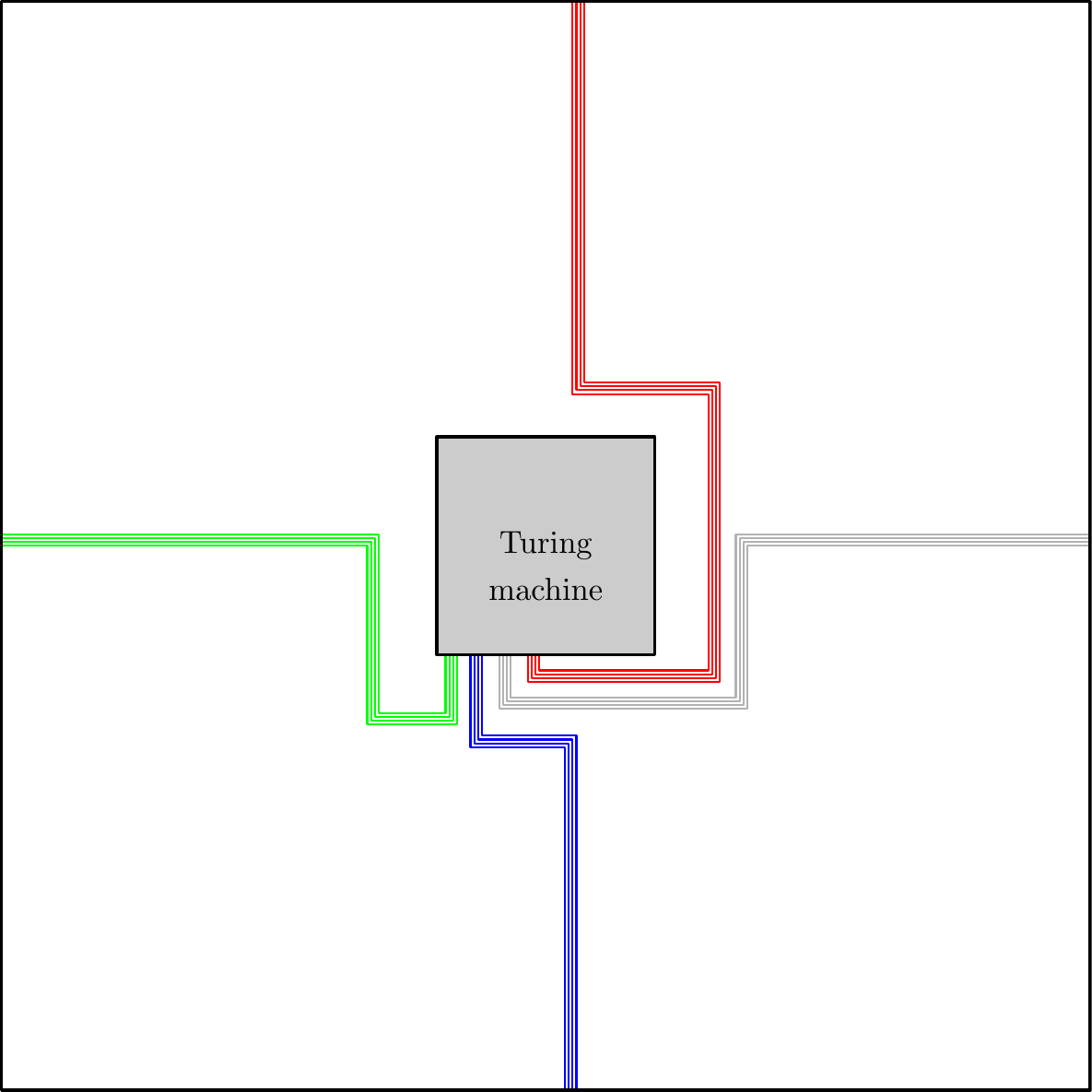}
\label{pic3}
\caption{}
\end{minipage}
\begin{minipage}[b]{0.45\linewidth}
\center
\includegraphics[scale=0.43]{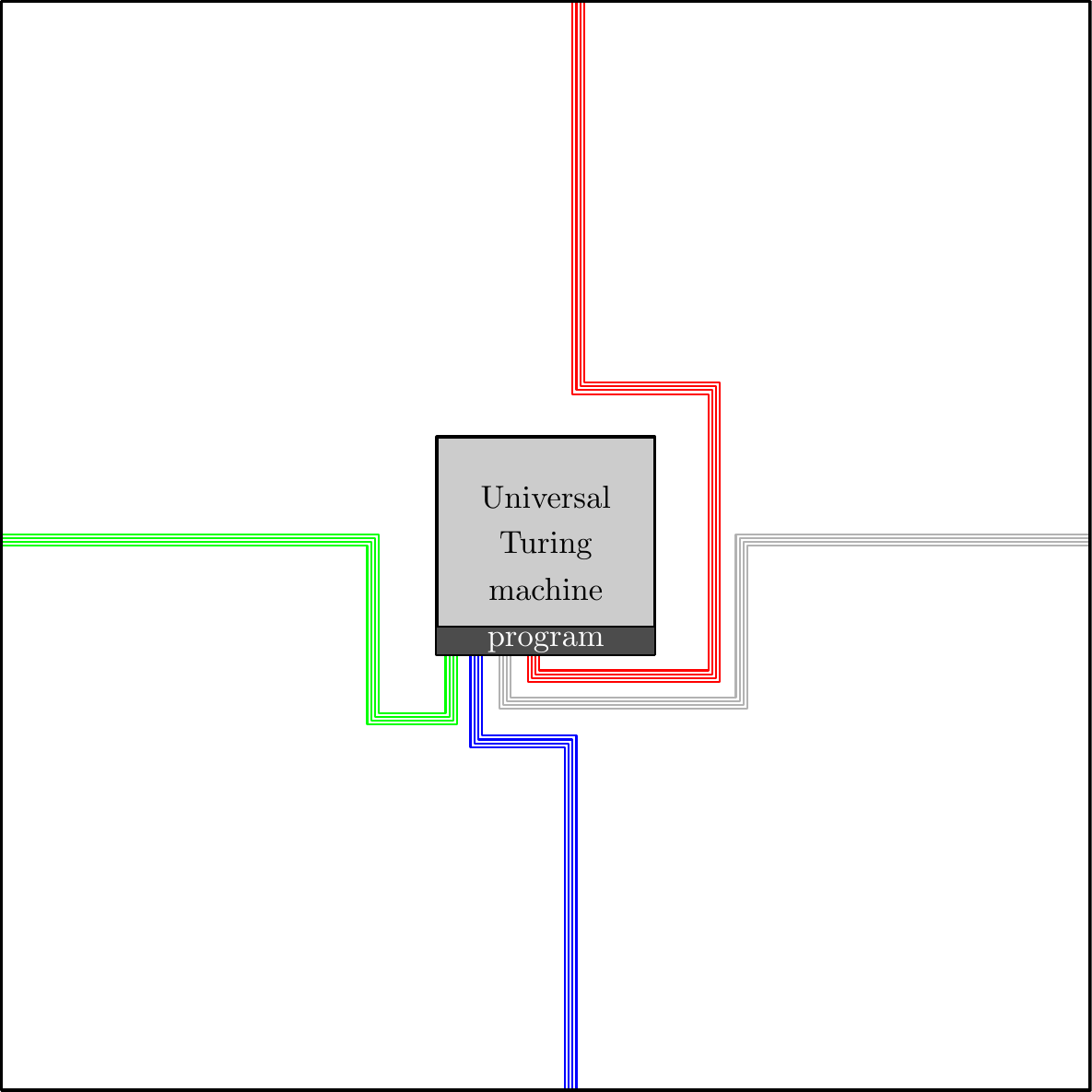}
\label{pic4}
\caption{}
\end{minipage}
\end{figure}

  To make this construction work, the size of macro-tile (the number $N$) should be large enough: first, we need enough space for $k$ bits to propagate, second, we need enough time (i.e., height) so all accepting computations of $\cal M$ terminate in time $m$ and on space $m$ (where the size of the computation zone $m$ cannot be greater than the size of a macro-tile).
 
In this construction the number of additional shades depends on the machine $\cal M$ (the more states it has, the more additional shades we need to simulate the computation in the space-time diagram). To avoid this dependency, we replace $\cal M$ by a fixed universal Turing machine $\cal U$ that runs a program simulating $\cal M$.  We may assume that the tape has an additional read-only layer. Each cell of this layer carries a bit that never  changes during the computation; these bits are used as a program for the universal machine. So in the computation zone the columns carry unchanged bits; the construction of a tile set  guarantees that these bits form the program for $\cal U$, and the computation zone of a macro-tile represents a view of an accepting computation for that program, see Fig.~4. In this way we get a tile set $\tau$ that has $O(N^2)$ tiles and implements $\rho$. (This construction works for all large enough $N$.)
%\begin{figure}
%\centering
%\includegraphics[scale=0.45]{pic4.pdf}
%\label{pic4}
%\caption{}
%\end{figure}

In the updated construction the tile set still depends on the program simulated in the computational zone. However, this dependency is essentially reduced: the simulated program (and, implicitly, the predicate $P$) affects only the rules for the tiles used in bottom line of the computational zone. The colors on the sides of all other tiles are universal and do not depend on the simulated tile set~$\tau$.

\subsection{A self-similar tile set:  implementing itself}

In the previous section we explained  how to implement a given tile set $\rho$ (represented as a program for the universal TM) by another tile set $\tau$ with large enough zoom factor $N$. Now we want $\tau$ be isomorphic to $\rho$. This can be done using a construction that follows Kleene's fixed-point theorem.
Note that most steps of the construction of  $\tau$ do not depend on the program for $\cal M$ (the coordinates of tiles that make the skeleton of a macro-tile, the information transfer along the wires, the propagation of unchanged program bits, and the space-time diagram for the universal machine in the computation zone). Let us fix these rules as part of $\rho$'s definition and set $k = 2 \log N + O(1)$, so that we can encode $O(N^2)$ colors by $k$ bits.  From this definition we  obtain a program $\pi$ for the TM  that checks that macro-tiles behave like $\tau$-tiles in this respect. We are almost done with the program $\pi$. The only remaining part of the rules for $\tau$ is the hardwired program. We need to guarantee that the computation zone in each macro-tile carries the very same program $\pi$. But since the program is written on the tape of the universal machine, it can be instructed to access its own bits and check that if a macro-tile belongs to the computation zone, this macro-tile carries the correct bit of the program.

It remains to explain the choice of $N$ and $m$ (note that the value of the zoom factor $N$ and the size of the computation zone $m$ are hardwired in the program). We need it to be large enough so the computation described above   (which deals with inputs of size $O(\log N)$) can fit in the computation zone. The computations are rather simple (polynomial in the input size, i.e., $O(\log N))$, so they easily fit in space and time bounded by $m=\poly(\log N)$.
This completes the construction of a self-similar aperiodic tile set.
 
Now it is not hard to verify that the constructed tile sets (1) allows a tiling of the plane, and (2) each tiling is self-similar. Applying Proposition~\ref{thm-folklore} we obtain the following proposition.
\begin{proposition}[R.~Berger]
There exists a tile set $\tau$ such that 
%\begin{enumerate}
%\item[(i)]  there exist $\tau$-tilings of the plane,
%\item[(ii)]  each $\tau$-tiling is  aperiodic.
%\end{enumerate}
 there exist $\tau$-tilings of the plane, and each $\tau$-tiling is  aperiodic.
\end{proposition}
 In the next section we will upgrade the basic construction of the fixed-point tile set. So far we should keep in mind that in such a tile set all tiles can be classified into three types:
 \begin{itemize}
 \item the ``skeleton'' tiles that keep no information except for their coordinates in a macro-tile; these tiles work as building blocks for our hierarchical structure;
 \item the ``wires'' that transmit the bits of macro-colors from the frontier of the macro-tile to the computation zone;
 \item the tiles of the computation zone (intended to simulate the space-time diagram of the Universal Turing machine). 
 \end{itemize}
 The same is true for macro-tiles, super-macro-tiles, etc.; i.e., each macro-tile is a  ``skeleton'' block, or a part of a ``wire'', or a cell in the computation zone in the macro-tile of higher rank.

 \section{Quasiperiodicity and aperiodicity}

Before we approach the main result, we prove a simpler statement; we show that there exists a tile set such that all tilings are both
\emph{quasiperiodic} and \emph{aperiodic}.
\begin{theorem}\label{thm2}
There exists a tile set \textup(a set of Wang tiles\textup) $\tau$ such that 
%%\begin{enumerate}
%%\item[(i)]  there exist $\tau$-tilings of the plane;
%%\item[(ii)]  each $\tau$-tiling is quasiperiodic; moreover, the set of $\tau$-tilings is minimal \textup(i.e., all $\tau$-tilings contain the same finite patterns\textup);
%%\item[(iii)]  each $\tau$-tiling is  aperiodic.
%%\end{enumerate}
(i)~there exist $\tau$-tilings of the plane;
(ii)~each $\tau$-tiling is quasiperiodic; moreover, the set of $\tau$-tilings is minimal \textup(i.e., all $\tau$-tilings contain the same finite patterns\textup);
(iii)~each $\tau$-tiling is  aperiodic.

\end{theorem}
This result was originally proven in \cite{alexis} (for a tile set $\tau$ constructed in \cite{ollinger}). 
%Theorem~\ref{thm2} is obviously weaker than Theorem~\ref{thm1} since every non-computable tiling is aperiodic.

 \subsection{Supplementary features: what else we can assume on the fixed-point tiling}
 
The general construction of a fixed-point tiling does not imply the property of quasiperiodicity. In fact, for tilings described above, each pattern that includes only ``skeleton'' tiles (or ``skeleton'' macro-tiles of some rank $k$) must appear infinitely often, in all homologous position inside all macro-tiles of higher rank. However, this is not the case for patterns that include tiles from the ``communication zone'' or the ``communication wires''. Informally, the problem is that  even a very small  pattern can involve the information relevant for a macro-tile of arbitrarily high rank. So we cannot guarantee that a similar pattern appears somewhere in the  neighborhood.  To overcome this difficulty we need some new idea and new technical tricks.
 %
%Fortunately, the construction of fixed point tilings is very flexible, and allows a  great variety  of generalisations.  
First of all, without essential modification of the construction we can enforce the following additional properties of a tiling:
\begin{itemize}
\item In each macro-tile, the size of the computation zone $m$ is much less than the size of the macro-tile $N$.  Technically, in what follows we will need to reserve free space in a macro-tile to insert $O(1)$ (some constant number) of copies of each $2\times 2$ pattern  from the computation zone (of this macro-tile). This requirement is easy to meet.
We may  assume that the size of a computation zone in a macro-tile of size $N\times N$ is only $m=\poly(\log N)$. 
\item We require that the tiling inside the computation zone   satisfies the property of $2\times2$-\emph{determinicity}:  if we know all  colors  on the borderline of a $2\times2$-pattern inside of the computation zone (i.e., a tuple of $8$ colors), then we can uniquely reconstruct the $4$ tiles of this pattern. Again, we do not need any new idea: this requirement is met if we simulate the space-time diagram of a Turing machine in a natural way.
\item The communication channels in a macro-tile   (the wires that transmit the information from the macro-color on the borderline of this macro-tile to the bottom line of its computation zone) must be isolated from each other. The distance between every two wires must be greater than $2$ from each other. That is, each $2\times 2$-pattern   can touch at most one communication wire.
\end{itemize} 
Also we will need a somewhat more essential modification of the construction. We discuss it in the next section.

\section{Proof of Theorem~\ref{thm2}}

To achieve the property of quasiperiodicity, we should guarantee  that every finite pattern that appears once in a tiling, must appear in each large enough square.  If a tile set $\tau$ is self-similar, then in every $\tau$-tiling  each finite pattern  can be covered by at most $4$  macro-tiles  (by a $2\times2$-pattern)  of an appropriate rank. Thus, to prove Theorem~\ref{thm2} it is enough to guarantee that each $2\times2$ group of macro-tiles (of each rank) that ever appears in a tiling, must appear  in eachl large enough squares in it. This property is not true for the tile set constructed above. As we noticed above, this is obviously true for a $2\times2$ pattern that involves only skeleton macro-tiles (we can find an identical pattern in the neighboring macro-tile of the appropriate rank); however, this property can be false for patterns that touch the communication wires or the computation zone.
To achieve the desired property we need to modify  the basic construction. To this end we implement in our construction one new feature.

\textbf{The new feature:} Notice that for each $2\times2$-window that touches the computation zone or the communication wires there exist only a bounded number $c$ of ways to tile them correctly (and make a correct tiling). This constant $c$  depends on the alphabet of the tape and the number of internal states of the Universal Turing machine. For each possible position of a $2\times2$-window in the computation zone or in the communication wires and for each possible filling of this window by tiles, we reserve a special $2\times2$-\emph{slot} in a macro-tile (somewhere far away from the computation zone and from all communication wires) and define the neighbors around this slot in such a way that only these specific $2\times 2$ patterns can patch it.  Note that the tiles around this  ``know'' their real coordinates in the bigger macro-tile, while the tiles inside the slot do not (they ``believe'' to be tiles in the computation zone, though they are in a ``slot'' outside of it). An example of such a slot is shown in Fig.~5. In Fig.~6 we show how these ``slots'' are placed in a macro-tile. This simple trick is the sharpest difference between this construction and the fixed-point tilings known before: now some tiles do  not ``know'' their real position in the ambient macro-tile.  
\begin{figure}
\centering
\includegraphics[scale=0.25]{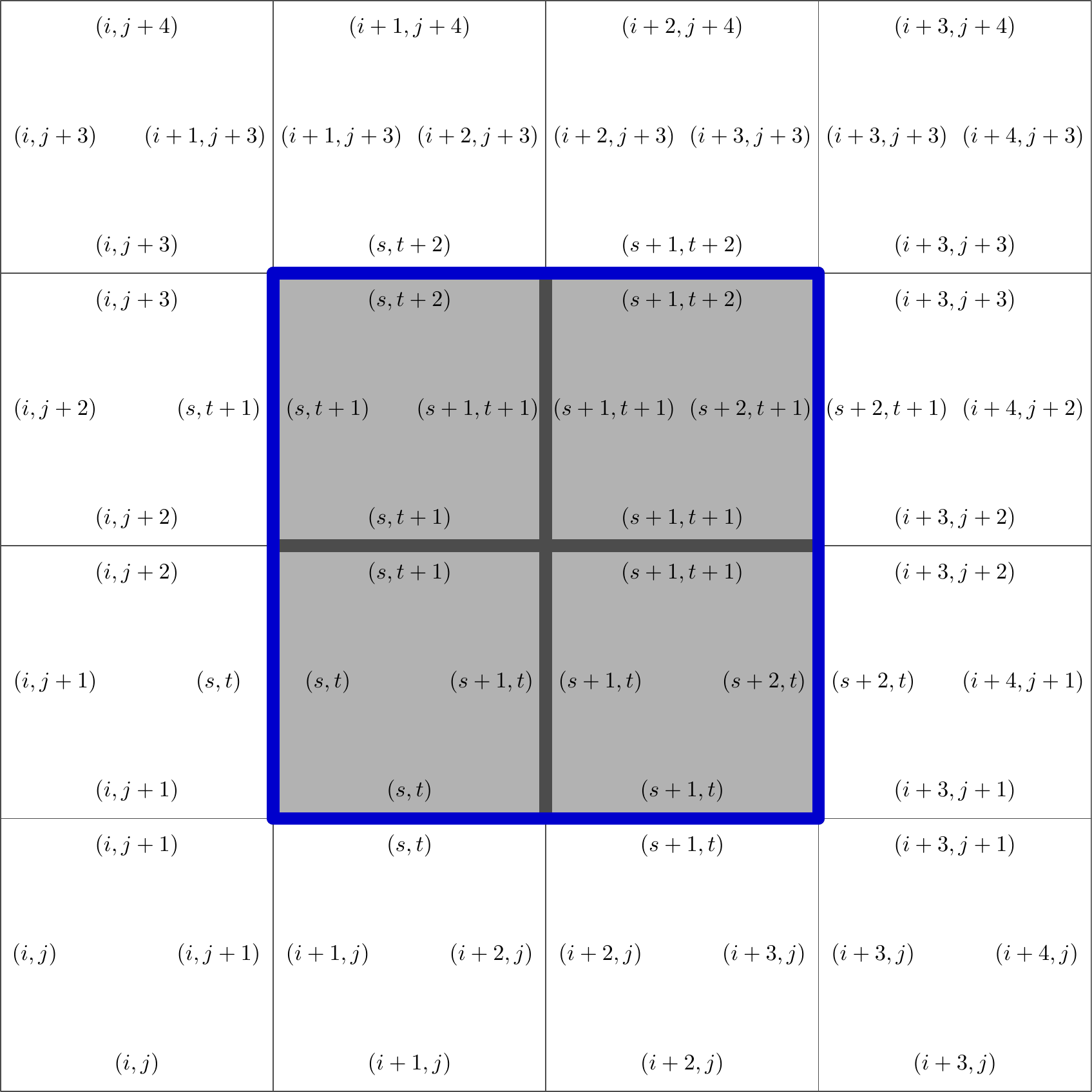}
\label{pic5}
\caption{A ring of 12 ``skeleton'' tiles (the white squares) makes a slot for a $2\times 2$-pattern of tiles from the computation zone (the grey squares). In the picture we show the ``coordinates'' encoded in the colors on the sides of each tile. 
The colors of the bold lines (the blue lines between white and grey tiles and the bold black lines between grey tiles) contain some information beyond coordinates --- these colors involve the bits used to simulate a space-time diagram of the universal Turing machine. (We do not show all the corresponding bits explicitly.) 
The ``real'' coordinates of the bottom-left corner of this slot are $(i+1,j+1)$, while the ``natural'' coordinates of the corresponding patterns (when it appears in the computation zone) are $(s,t)$. 
}
\end{figure}
\begin{figure}
\centering
\includegraphics[scale=0.5]{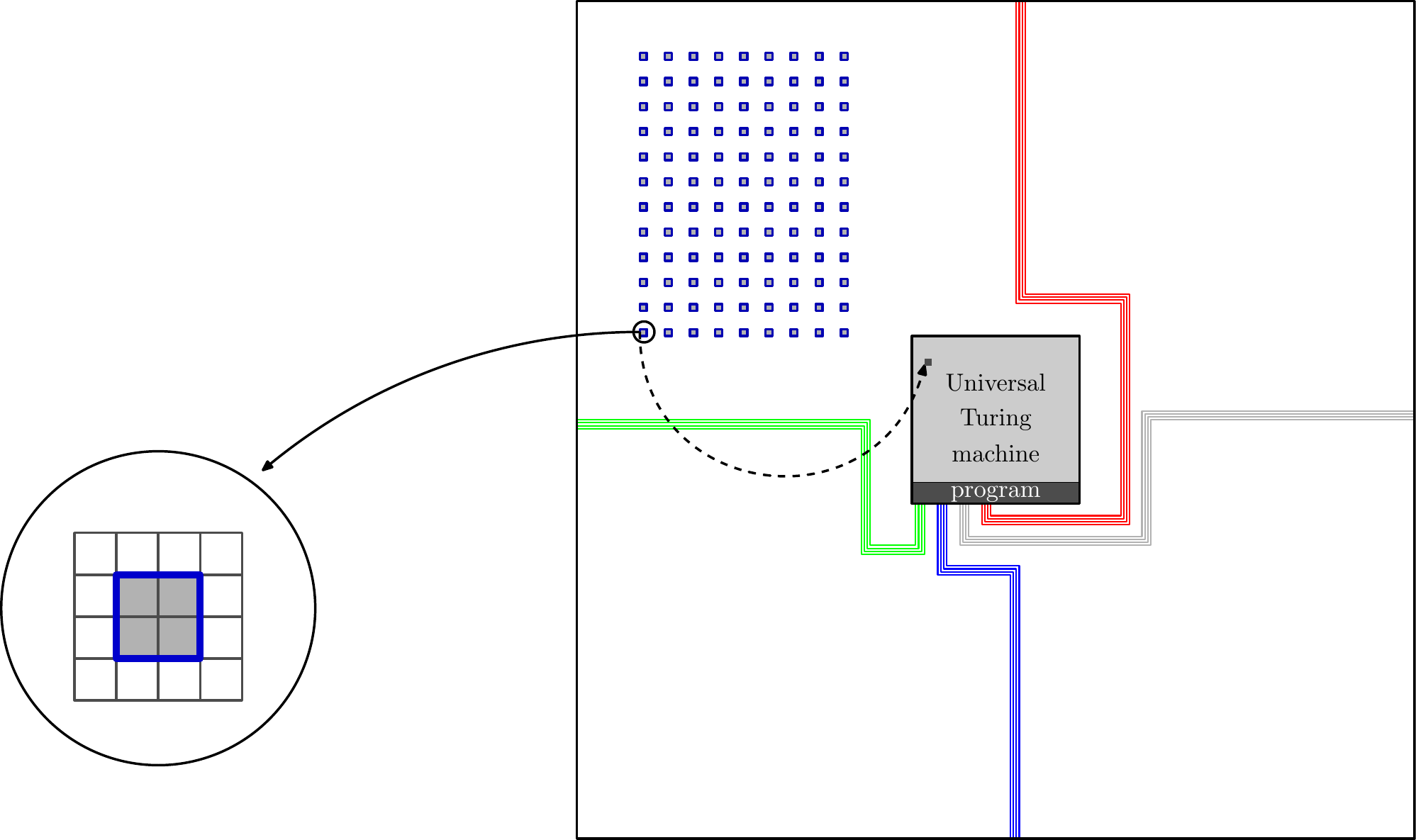}
\label{pic6}
\caption{The array of ``slots'' (with patterns from the computation zone) embedded in a macro-tile.}
\end{figure}

Here we use (a) the property of  $2\times2$-determinicity of the computation zone (there is a unique way to put tiles in the ``slot''), and (b) the fact that we have enough room to put in a macro-tile the slots for all $2\times 2$-patters that can appear in the computation zone or along the communication wires. (Here we use the fact that the size of the computational zone $m\times m$ and the lengths of all communication wires $O(1)\times O(N)$ in a macro-tile are much less than the total area of a macro-tile  $N\times N$.) This feature guarantees that each $2\times2$ pattern from the computational zone appears at least once in each macro-tile (such a pattern appears once in each macro-tile  in the introduced ``slots'' and possibly once again in the computation zone of this macro-tile).

We choose the positions of the ``slots'' in the macro-tile so that coordinates can be computed by a short program in time polynomial in $\log N$. We require that the positions of all slots are disjoint, and they do not touch each other. This precaution is needed to guarantee that the tiles used in the slots do not damage the general structure of the macro-tiles.

For a tile set with a new feature, every tiling enjoys a new property: every $2\times 2$-pattern of tiles touching the computation zone or a communication wire, must appear at least once in \emph{each} macro-tile (hence, this pattern must appear in each large enough square). Of course, the property of self-similarity implies that similar statements hold for  $2\times 2$-pattern of macro-tiles of each ranks $k$.

Thus, we proved that every $N^k\times N^k$ pattern that appear in a $\tau$-tiling, must appear in each large enough square in this tiling; moreover,  this pattern must appear in each large enough square in \emph{all} $\tau$-tilings. Hence, the constructed tile set satisfies the requirements of Theorem~\ref{thm2}. 

\section{From aperiodicity to non-computability}

To prove Theorem~\ref{thm1}, we need a slightly more sophisticated construction. We need a self-similar tiling with \emph{variable zoom factor}, see
 \cite{drs} for details. In this version of the construction the size of a macro-tile of rank $r$ is equal to $N_r\times N_r$, for some suitable sequence of zooms $N_r$, $r=1,2,\ldots$ We may assume that $N_r = Cr$ for some constant $C$.   Now each macro-tile of rank $r$ must ``know'' its own rank (that is, the binary representation of $r$ is written on the tape of the Turing machine simulated on the computation zone). This information is used by a macro-tile to simulate the next rank macro-tiles properly.  The size of the computational zone $m_r$ should also grow as a  function of rank $r$ (easily computable from $r$); again, we may assume that $m_r = \poly(\log N_r)$.

Also we may require that all macro-tiles of rank $r$ contain in their computational zone the prefix (e.g., of length $\lceil \log r \rceil $) of some infinite sequence
 $X=x_0x_1x_2\ldots$
The bits of this prefix are propagated by wires to the neighboring macro-tiles, so all macro-tiles of the same rank contain the same bits $x_0x_1\ldots$  The usual self-simulation guarantees that the bits of $X$ embedded into a macro-tile of rank $r+1$ extends the prefix embedded in a macro-tile of rank $r$. Since the size of the computational zone increases as a function of rank $r$, the entire tiling of the plane involves an infinite sequence of bits $X$.

The construction becomes interesting if we can enforce some special properties of the embedded sequence $X$. For example, we can guarantee that it is not computable. Indeed, let us make the machine in the computation zone do some new job: let it enumerate two non-separable enumerable sets (on each level $r$ we run the simulation for the  number of steps that fits the  computation zone available in a macro-tile of rank $r$). Then we can require that $X$ is a separator between these two sets, and in each level,  the machine verifies that the (partially) enumerated sets are indeed separated by the given prefix of $X$.
Combining all ingredients together, we obtain a tile set $\tau$, which is self-similar in a generalised sense (with a variable zoom factor), with two nontrivial properties: all $\tau$-tilings are non-computable and quasiperiodic. Thus,  we proved Theorem~\ref{thm1}. 

\noindent
\emph{A technical remark}:
Notice that in this construction we cannot control precisely the sequence $X$ embedded in the tiling (we can specify  the two non-separable enumerable sets that are ``enumerated'' in a tiling, but we cannot define uniquely the separator $X$ between them). Thus, our tile set accepts tilings corresponding to infinitely many sequences $X$, and it is not minimal. This is not surprising:  if an effectively closed subshift contains no computable points, then it cannot be minimal (see, e.g., \cite{hochman-2009}).
 In contrast to the proof of Theorem~\ref{thm1-bis},
in the construction used in this section we cannot claim that there exist only $O(1)$ ways to fill 
by a quadruple of macro-tiles of rank $k$
a slot of size $2N_k\times2N_k$  placed somewhere in a macro-tiles of rank $(k+1)$. Indeed, these macro-tiles involve a prefix of  $X$, and there exist potentially many different sequences $X$. However, once $X$ is fixed, there rest only a constant number of $2\times2$ macro-tiles of rank $k$ that fit the given position in the next level macro-tile. This observation allows to reuse the ``new feature'' from the proof of Theorem~\ref{thm2}.

%\medskip

%\emph{Remark:} Let $T$ be some $\tau$-tiling, and $X$ be the infinite non-computable sequence embedded in this tiling. Then all $\tau$-tilings $T'$ equivalent to $T$ (i.e., tilings that contain exactly the same finite patterns) contain the same $X$. Thus, if the embedded sequence $X$  belongs to a high enough Turing degree (e.g., it is not low, not hyperimmune-free, etc.), then $T'$ must also  belong to a high enough Turing degree (respectively, not low, not hyperimmune-free, etc.).  However we cannot guarantee that all $\tau$-tilings are not low or not hyperimmune-free (this is impossible because of the low basis theorems).

%\smallskip

With essentially the same technique we can prove Theorem~\ref{thm1-bis}. We employ again the idea of embedding of an infinite sequence $X$ in a tiling. Technically, we require that all macro-tiles of rank $k$ should involve on their computational zone the same finite sequence of $\log k$ bits, which is understood as a prefix of $X$; we guarantee that the prefix embedded in macro-tiles of rank $(k+1)$ is compatible with the prefix available to the macro-tiles of rank $k$. Further, since $\cal A$ is in $\mathrm \Pi_1^0$, we can enumerate the (potentially infinite) list of patterns that should not appear in $X$. On each level, the macro-tiles allocate some part of  the available space and time (limited by the size of the computational zone available on this level) to run the enumeration of $\cal A$; every time a new element of $\cal A$ is enumerated,  the algorithm (simulated in the computational zone) verifies that the found forbidden pattern does not appear in the prefix of $X$ accessible to macro-tiles of this level. Since the computational zone in a macro-tile of rank $k$ becomes bigger and bigger as $k$ increases, the enumeration extends longer and longer. Thus, a sequence $X$ can be embedded in an infinite tiling, if and only if this sequence does not contain any forbidden pattern (i.e., this $X$ belongs to $\cal A$).

What are the Turing degrees of tilings in the described tile set? For our tile set, every tiling is %uniquely 
defined by three infinite parameters: the sequence of bits $X$ embedded in this tiling, and two sequences of integers $\sigma_h, \sigma_v$ that specifies the shifts (the vertical and the horizontal ones) of macro-tiles of each level relative to the origin of the plane. This information is enough to reconstruct the tiling. Indeed, $\sigma_h$ and $\sigma_v$ define the hierarchical  structure of the macro-tiles: on each level $k$ we should split the macro-tiles of the previous rank into blocks of size $N_k\times N_k$ ($k$-level macro-tiles), and there are $N_k^2$ ways to choose the grid of horizontal and vertical lines that define this splitting. And the content of the computational zones of all macro-tile is defined by the prefixes of $X$.
%All the tiling is uniquely defined if we know the triple $(X,\sigma_h, \sigma_v)$. 
Conversely, given a tiling as an oracle, we can (computably) extract from it the digits of the sequences $X$, $\sigma_h$, and $\sigma_v$.
% --- given these sequences and an $X\in{\cal A}$ we can always construct the corresponding tiling.  
It remains to notice that $\sigma_h$ and $\sigma_v$ can be absolutely arbitrarily.
Thus, the Turing degree of a tiling is the Turing degree of  $(X,\sigma_h, \sigma_v)$, which can be arbitrary degree not less than $X$. That is, the set of degrees of tilings is exactly the upper closure of ${\cal A}$. %, i.e., the set of all $Y$ that are not less than some $X\in{\cal A}$. 
So we get the statement of  Theorem~\ref{thm1-bis}.

\smallskip

\emph{Acknowledgements:} We thank Laurent Bienvenu and Emmanuel Jeandel for many prolific discussions. We are also very grateful to the three anonymous referees of MFCS 2015 for exceptionally detailed and instructive comments.

\end{document}